\documentclass[pss]{wiley2sp} 
 \pdfoutput=1
                        \usepackage{amsmath}
\usepackage{amssymb}
\usepackage{bm}              
\usepackage{w-greek}         
\usepackage[multidot]{grffile}

\usepackage{ushort}
\usepackage{bbm}
\usepackage{rotating}
\usepackage{pgf}
\newcommand{\dul}[1]{\ensuremath{\ushortd{#1}}}

\newcommand{\ket}[1]{|#1\rangle}

\begin{document}

\title{Mott-Hubbard transition in V$_2$O$_3$ revisited}

\titlerunning{Mott-Hubbard transition in V$_2$O$_3$ revisited} 


\author{P. Hansmann\textsuperscript{\textsf{\bfseries 1,2}}, A. Toschi\textsuperscript{\textsf{\bfseries 1}}, G. Sangiovanni\textsuperscript{\textsf{\bfseries 1,3}}, T. Saha-Dasgupta\textsuperscript{\textsf{\bfseries 4}}, S. Lupi\textsuperscript{\textsf{\bfseries 5}}, M. Marsi\textsuperscript{\textsf{\bfseries 6}}, K. Held\textsuperscript{\textsf{\bfseries 1}}}
\authorrunning{P. Hansmann et al.}
\institute{
  \textsuperscript{1}\, Institute of Solid State Physics, Vienna University of Technology,
 1040 Vienna, Austria\\
  \textsuperscript{2}\, Centre de Physique Th\'eorique, \'Ecole Polytechnique, CNRS, 91
128 Palaiseau, France\\
    \textsuperscript{3}\, Institut f\"ur Theoretische Physik und Astrophysik, Universit\"at W
\"urzburg, Am Hubland, D-97074 W\"urzburg, Germany\\
  \textsuperscript{4}\, S.N. Bose National Centre for Basic Sciences,
Kolkata 700098, India\\
  \textsuperscript{5}\,CNR-IOM and Dipartimento di Fisica, Universit\`a di 
Roma "La Sapienza",
Piazzale A. Moro 2, I-00185 Roma, Italy\\
  \textsuperscript{6}\,Laboratoire de Physique des Solides, CNRS-UMR 8502, 
Universit\'{e} Paris-Sud, F-91405 Orsay, France
}

\received{XXXX, revised XXXX, accepted XXXX} 
by the publisher
\published{XXXX} 


\abstract{
The isostructural metal-insulator transition in Cr-doped V$_2$O$_3$ is  the
textbook example of a Mott-Hubbard transition between a paramagnetic metal and a paramagnetic  insulator. We review recent theoretical calculations as well as experimental findings which shed new light on this famous transition. In particular, the old paradigm of a doping-pressure equivalence does not hold, and there is a microscale phase separation for Cr-doped V$_2$O$_3$.
}
\maketitle                   

\section{The story so far}
\begin{figure}
\sidecaption
    \includegraphics{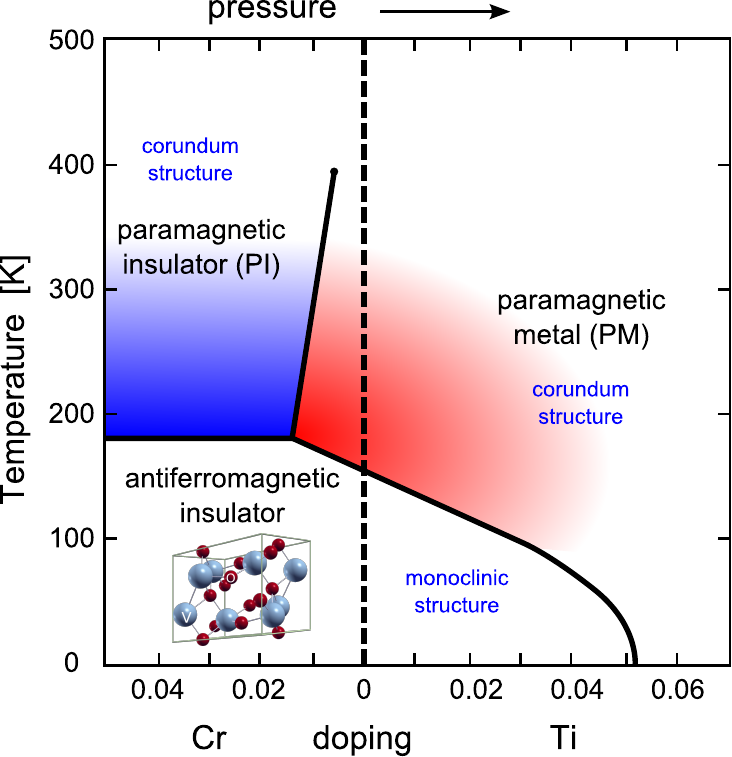}
\caption{(Color online) Temperature vs. doping phase diagramm of V$_2$O$_3$  according to \cite{mcwhan73}. In the PM and PI phase the compound crystallizes in the corundum structure, whereas the low temperature AF phase shows a monoclinic lattice structure. The double x-axis as pressure and doping reflects the paradigm of pressure-doping equivalence  \cite{mcwhan70}.
 The monoclinic and corundum structure unit cells are shown as small insets. \label{V2O3_phase}}
\end{figure}

At first, let us summarize the basic facts and review
some of the former theoretical work that has been put forward, thereby also
defining the necessary terms.
In Fig.~\ref{V2O3_phase} we show the phase diagram of V$_2$O$_3$ \cite{mcwhan70}
spanned in the temperature--doping space displaying three phases:
At ambient conditions V$_2$O$_3$ is a paramagnetic metal (PM) and crystallizes in the
corundum structure with four vanadium atoms in the primitive
unit cell, see inset of 
Fig.~\ref{V2O3_phase}. It can be seen that respectively two
vanadium atoms form ``pairs'' which are oriented along the
crystallographic c--axis. Upon
cooling below $\sim$150K, a peculiar antiferromagnetic  order (AF) sets in and the
system becomes insulating, accompanied by a monoclinic structural
distortion. On the other hand, the system can be tuned by doping
with chromium or titanium or the application of external
pressure. In this respect, the ``common wisdom'' has been established \cite{mcwhan70} that doping and pressure can be seen
as equivalent routes through the phase diagram. As we will see later, however, the pressure/doping equivalence scheme is inconsistent with recent
experimental measurements of the optical conductivity and x--ray
absorption. Above the N\'{e}el, temperature, the corundum crystal structure remains unchanged as a function of pressure or
doping. Nontheless, upon Cr doping a first order isostructural metal--to--insulator
(MIT) transition takes place (see Fig.~\ref{V2O3_phase}) which
evoked several theoretical attempts to describe this MIT as a genuine
Mott--Hubbard transition. While the MIT is associated to changes in
the lattice structure and the atomic positions
\cite{mcwhan69,robinson75}, it is important to notice that x--ray
diffraction showed that for a given temperature the structure within
one phase \emph{does not change upon doping} \cite{robinson75}.
It was later also observed by Park \emph{et al.} \cite{park00} with vanadium
$L$--edge x--ray absorption spectroscopy that this holds
also the electronic configuration of the system in terms of the orbital occupation (see Table 1 of
\cite{park00}). Therefore we shall adopt the nomenclature of Robinson
\cite{robinson75} and refer to the lattice structure of the paramagnetic metallic (PM) 
and insulating (PI) phase \emph{at ambient pressure} as $\alpha$-- and
$\beta$--phase respectively.

\begin{figure}
  \begin{center}
\includegraphics[width=8cm]{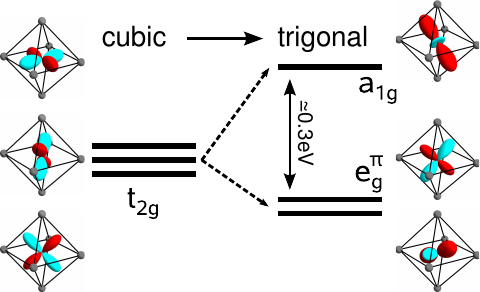}
  \end{center}
    \caption{(Color online)  Level splitting of the vanadium $t_{2g}$--states: due to
      the trigonal distortion the $t_{2g}$--states split up in an
      $e^\pi_g$ doublet and an $a_{1g}$ singlet at a higher energy. From LDA results we
      estimate approximately 0.3 eV  for this splitting. The plotted orbitals are
      spherical harmonic functions which display the
      symmetry of the states. These plots nicely compare to the
      NMTO Wannier function plots of Saha--Dasgupta \emph{et al.} \cite{tanusri09}  (reprinted with permission from \cite{hansmann12a}
 Copyrigth (2012) by the American Physical Society).}
    \label{V2O3_lvl}
\end{figure}

The electronic configuration of atomic vanadium is [Ar]3d$^3$4s$^2$,
which means, that in the three--valent oxidation state we find a
3d$^2$ configuration realized. In the corundum type structure the
vanadium atoms are coordinated by oxygen ligands in a trigonally
distorted octahedral fashion (inset of  Fig.~\ref{V2O3_phase}). Hence,
the cubic part of the crystal field splits the d--states into the
lower lying $t_{2g}$ and the higher lying $e_g$ states. The trigonal
distortion\footnote{Let us, already in this introduction, remark that
  the actual crystal field breaks one significant point symmetry on the
  vanadium sites, namely inversion in the c--direction. This is
  related to the different distances of the neighboring vanadium atoms
  along the c--axis, see Fig.\ \ref{V2O3_phase}. While this effect is
  negligible for most of the discussion, it will be
  of great importance for the selection rules of the polarization
  dependent XAS results later on.} acts like a compression along one
of the three--fold axes of the octahedron (i.e. squeezing two opposite
sides together). As a result the degeneracy of the lower lying
$t_{2g}$ states is lifted and they are split into a single $a_{1g}$
and the twofold degenerate $e_{g}^\pi$ states. This level splitting,
together with a plot of the respective angular part of the (atomic) wave
function, is sketched in Fig.~\ref{V2O3_lvl}. To indicate the difference to the $t_{2g}$
states, the higher lying cubic $e_g$ states (which are not split by
the trigonal distortion) get an additional index $e_g^\sigma$ in order
to distinguish them from the $e_g^\pi$. The $\sigma$
accounts for their orientation towards the ligands, with which
they form $\sigma$ bonds.\\ 
Since the $e_g^\sigma$ are pushed up in energy by the crystal field, the two
vanadium d--electrons populate the three $t_{2g}$ levels. One of the crucial aspects concerning the
understanding of the MIT is the specific occupation of these $t_{2g}$
states. In an early work, Castellani et al. \cite{castellani78A,castellani78B,castellani78C} assumed a
strong hybridization of the V--V pairs oriented parallel to the
rombohedral c--axis, resulting in a strong bonding and antibonding
splitting of the $a_{1g}$ states. In this case, with the bonding
states filled there would be one electron remaining in the twofold
degenerate $e_g^\pi$ states and the compound could be described by a
quarter filled $S=1/2$ Hubbard model. However, later experimental
evidence demonstrated \cite{moon70,paolasini99,dimateo02,park00} that
the ground state of the system is more complicated and should rather be
described as a $S=1$ state consisting of a mixture of $a_{1g}$ and
$e_{g}^\pi$ electrons.\\
Moreover, it is precisely the coefficients in the linear
combination of $a_{1g}$ and $e_{g}^\pi$ for the ground state which
allow for a quantitative distinction of the PM, PI, and AF
phases. The XAS vanadium $L$--edge study of Park \emph{et al.}
explored the phase diagram by means of temperature and doping and
summarized the respective ratio of  $e_{g}^\pi e_{g}^\pi$ to
$a_{1g}e_{g}^\pi$ occupations as \cite{park00}:
1:1 (PM), 3:2 (PI), and 2:1 (AF).
 As we mentioned earlier, their results turned out to
be consistent with the x--ray diffraction data for the lattice of Robinson
\cite{robinson75}, and showed that, within the PM $\alpha$-- and the
PI $\beta$-- phase, there is essentially no change in the ground state composition
for different doping levels. One main new result, which will be discussed
later, is, that this is not true for the
pressurized metallic phase.

\begin{figure}
  \sidecaption
      \includegraphics{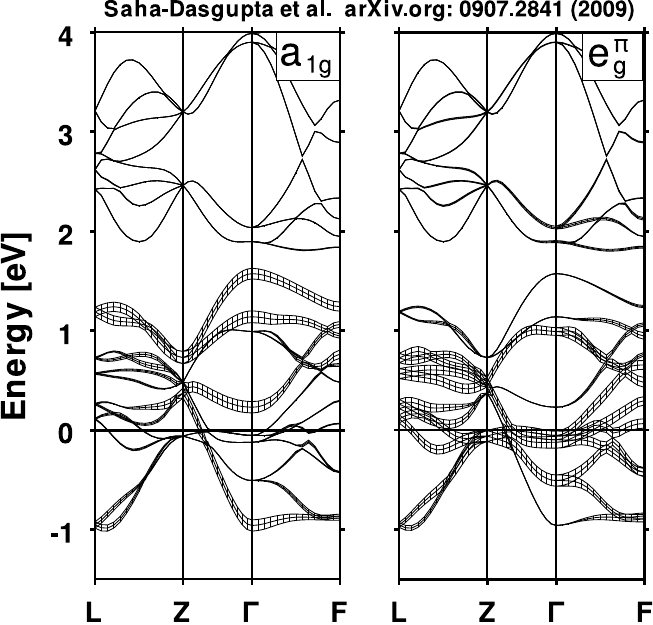}
  \caption{ V$_2$O$_3$ bandstructure from LDA with ``fat bands'' indicating the orbital
      character. Left: $a_{1g}$; right: $e_g^{\pi}$ orbital character. (taken from Saha--Dasgupta \emph{et al.}\cite{tanusri09}) \label{V2O3_tanusribands}}%

\end{figure}

{\bf LDA calculations } 
First \emph{ab initio} band structure calculations in the local density approximation (LDA) for V$_2$O$_3$ were
performed by Mattheiss \cite{mattheiss94}. Not surprisingly the results 
neither captured the insulating character of the Cr doped PI phase nor
the signatures of the strongly correlated character of the undoped PM
phase (for example the photoemission spectral weight identified with
the lower Hubbard band). Yet, even at the LDA level,  some
valuable information can be obtained. In Fig.~\ref{V2O3_tanusribands},
a plot of the LDA band structure is shown in the rombohedral
representation taken from Saha--Dasgupta \emph{et al.}\cite{tanusri09} with Fermi level
$\varepsilon_{\text{F}}=0$. In the two panels the respective $a_{1g}$ and
$e_g^\pi$ character is indicated by the width of the lines by means of
the so called ``fat band'' representation.\\
First of all it can be seen in Fig.\ \ref{V2O3_tanusribands} that the $t_{2g}$ part is
nicely separated from the rest of the bands: Genuine oxygen $p$--bands are
lower lying than the displayed energy range and the $e_g^\sigma$ can
be identified as the bands at $\approx 2$ eV to $4$ eV.\\
Let us now turn to the $a_{1g}$
bands (left panel Fig.~\ref{V2O3_tanusribands}). The previously
mentioned bonding--antibonding splitting due to the V--V pairs can be
seen at the $\Gamma$--point to be $\gtrsim 2$ eV, where the $a_{1g}$--character is
pure. The strongest dispersion is observed along the $\Gamma$--$Z$ direction where
the main contribution stems from $a_{1g}$--$a_{1g}$
hopping. However, the dispersion of the $a_{1g}$--bands along the
other directions is also not small which is a consequence of
$a_{1g}$--$e_g^\pi$ hybridization. While the $a_{1g}$ and $e_g^\pi$ states are 
orthogonal eigenstates \emph{locally} in a trigonal crystal field they
still hybridize in a \emph{non--local} way, i.e., there is intersite
$a_{1g}$--$e_g^\pi$ hopping also in the ab--plane. As it was
remarked by Elfimof \emph{et al.} \cite{elfimov03} these kinds of hopping are
important for the shape of the $a_{1g}$ states. With the help of the
fat bands in Fig.~\ref{V2O3_tanusribands}, we compare the
$a_{1g}$ and $e_g^\pi$--character: We clearly see that only at the
high symmetry points in the Brillouin zone the non--local
$a_{1g}$--$e_g^\pi$ hybridization is zero. 
In fact, this LDA result contradicts also theoretically the validity of
quarter filled $e_g^\pi$ states and the $S=1/2$ scenario.

{\bf Previous LDA+DMFT studies } 
By means of
dynamical mean field theory (DMFT) \cite{Georges96a} it is possible to include local electronic correlations, which trigger a Mott-Hubbard transition. 
Starting from the LDA results, first LDA+DMFT calculations
were performed and compared to photoemission and XAS experiments by
Held \emph{et al.} \cite{held01B} and Keller \emph{et al.}
\cite{keller04}. Later Poteryaev \emph{et al.} \cite{poteryaev07}
performed new LDA+DMFT calculations employing  a downfolded NMTO
$t_{2g}$ Wannier functions Hamiltonian provided by Saha--Dasgupta \emph{et al.} \cite{tanusri09} instead of the density of states \cite{held01B,keller04}.
Both kind of LDA+DMFT calculations capture the Mott-Hubbard transition, and well agree with (even predicted) the  photoemission spectroscopy (PES) measurements. Actually, the PES spectra
at the time of the first calculations  \cite{held01B} still did not show a pronounced quasiparticle peak  \cite{Schramme00}. Only after improving the bulk sensitivity by using high-energy photons from a synchrotron source
\cite{Mo02,Mo06a}, experiments agreed also in this respect with LDA+DMFT.

Extending the
work of Poteryaev \emph{et al.} \cite{poteryaev07}, Tomczak
\cite{tomczakthesis} and Tomczak and Biermann \cite{tomczak09}
discussed the optical conductivity of the compound
introducing corrections for the calculation of the Fermi--velocities
associated with the non--monoatomic basis of V$_2$O$_3$ -- an important
issue also for this work which will be discussed in the next
section. The last work that should be mentioned is the joint
experimental/theory paper by Baldassare \emph{et al.}
\cite{baldassarre08} in which the authors show that the slight change of
the lattice parameters due to temperature, drive the system
into the crossover regime between metal and insulator.
Their results underline how sensitive strongly correlated systems are
with respect to the change of external parameter -- even more so in 
the vicinity of a correlation driven Mott transition. For a comparison
to other vanadium oxides, see \cite{Perucchi09b}.

The key interest of the more recent LDA+DMFT calculations, which will be discussed in the
following, is to shed new light on the actual ground state of
V$_2$O$_3$ at different points in the phase diagram
Fig.~\ref{V2O3_phase}. Special attention is paid to the insulating and
metallic phase of the 1.1\% Cr--doped sample in the vicinity of the
MIT as well as to the comparison between the metallic phase of the undoped
sample at ambient conditions and the Cr--doped sample under external pressure.

In the following, we discuss some aspects and details
of the LDA+DMFT calculation in  
Section \ref{Sec:LDADMFT}. Theoretical and experimental 
results for  the optical conductivity are
presented in Section 
\ref{Sec:optcond}, those for the photoemssion microscopy 
in Section \ref{Sec:photomicro}, and those for the X-ray absorption spectroscopy
in Section \ref{Sec:xray}. Finally, Section   \ref{Sec:conclusion} provides a summary and conclusion.

\section{LDA+DMFT implementation}
\label{Sec:LDADMFT}

\subsection{Low energy $t_{2g}$ NMTO Hamiltonian}
\label{V2O3_model}
The first step of the LDA+DMFT calculations is the derivation of the
Hamiltonian for the low-energy $t_{2g}$ orbitals  from the bandstructure calculation via
NMTO downfolding or Wannier projections. The Hamiltonian is
constructed to capture the relevant degrees of freedom of the system for low
energy scales on a reduced basis set. In the case of V$_2$O$_3$ we
used a model obtained by the NMTO method, with which the full LDA Hamiltonian 
was downfolded on the $t_{2g}$ sub--space around the Fermi energy. As
described above (see Fig.~\ref{V2O3_lvl}), the $t_{2g}$ states
are decomposed into a single $a_{1g}$ and two degenerate $e_g^\pi$
states. However, if we look closely at the bandstructure in
Fig.~\ref{V2O3_tanusribands} we find twelve $t_{2g}$ bands instead of
three. The reason for this is simply that there are four vanadium
atoms in the primitive unit cell which means, that we obtain a 12 by 12
Hamiltonian as a function of $\vec{k}$ for V$_{2}$O$_3$ from the
downfolding. (For a detailed discussion of the 
downfolding procedure of the V$_2$O$_3$ model see Saha--Dasgupta
\emph{et al.} \cite{tanusri09}). Yet, although the LDA Hamiltonian is
a twelve--band
dispersion matrix, the actual DMFT calculation can be performed with
no more effort than a three band calculation. The reason for this is
simply that all four vanadium atoms in the unit cell are located on 
equivalent sites which means that they are related to one another by
symmetry transformations. In other words, each of the four vanadium
atoms experiences the same 
environment and, hence, has the same \emph{local} eigenstates. As a
consequence, the ${\mathbf k}$--integrated local Green function can be written in a basis in which we 
obtain four equal diagonal blocks with respect to the site index. The orbital labels $a_{1g}$ and $e_g^\pi$ are good quantum numbers
\emph{locally}. Such a local basis set is a necessary condition for
the formulation of the local interaction parameter U and a correct definition of the
\emph{local} DMFT self energy.\\

\subsection{DMFT Green function and self energy}
We explicitly write the local Green function as:\\
\begin{equation}\label{V2O3_gloc}
  \begin{split}
  \dul{G}^{\text{loc.}}(\omega)& =\frac{1}{V_{\text{BZ}}}\int_{\text{BZ}}d^3k\frac{1}{(\omega+\mu)\dul{\mathbbm{1}}-\dul{\varepsilon}^{\text{LDA}}(\vec{k})-\dul{\Sigma}(\omega)}\\[0.4cm]
  &=\begin{pmatrix}
    \dul{G}^{\mathbf{I}}            &        & \dul{G}_{\text{hyb.}}\neq\dul{0}\\
                                   & \ddots &                                \\
    \dul{G}_{\text{hyb.}}\neq\dul{0}&         &\dul{G}^{\mathbf{IV}}            \\
  \end{pmatrix}
\end{split}
\end{equation}\\
where the roman numerals serve as a site index, and, as mentioned\\
\begin{equation}
 \dul{G}^{\mathbf{i}}=
  \begin{pmatrix}
    G_{a_{1g}}^{\text{loc.}}&0           &0          \\ 
    0        &G_{e_{g}^\pi}^{\text{loc.}}&0           \\
    0        &0           &G_{e_{g}^\pi}^{\text{loc.}}\\
  \end{pmatrix}; \;\;\;\;\;
  \mathbf{i}=\mathbf{I},\cdots,\mathbf{IV}\\
\end{equation}\\
the diagonal blocks are equal for each site i. Hence, in order to
calculate the local DMFT self energy, which clearly has to be
the same for all four (locally equivalent) sites, we have to  project out only the first
diagonal block after the $\vec{k}$--integration, and proceed with the
calculation of the DMFT self energy in the usual way. 
The resulting self energy is a diagonal 3 by 3 matrix
$\dul{\Sigma}^{\mathbf{I}}(\omega)$ and is used in the
next iteration to construct the full 12 by 12 diagonal matrix
$\dul{\Sigma}(\omega)$ taking the equality of the four vanadium sites
into account\\
\begin{equation}
\dul{\Sigma}(\omega)=\begin{pmatrix}
      \dul{\Sigma}^{\mathbf{I}}(\omega)&0                        &0&0          \\ 
      0        & \dul{\Sigma}^{\mathbf{I}}(\omega)           &0     &0      \\
      0        &0           & \dul{\Sigma}^{\mathbf{I}}(\omega)&0\\
      0        &0           &0                      & \dul{\Sigma}^{\mathbf{I}}(\omega)\\
    \end{pmatrix}
\end{equation}\\
This full self energy then enters equation \eqref{V2O3_gloc} for the calculation of
the next local Green function.  Furthermore, it is important to strongly stress
at this point that we do not make additional approximations with the
procedure described above. From the DMFT point of
view, i.e., the local perspective, the V$_2$O$_3$ calculation is de facto
just a three orbital problem.

Besides the additional step of projecting out the local part of the
full Green function the LDA+DMFT calculation of V$_2$O$_3$ is straight
forward, see \cite{LDADMFT,Kotliar06,heldreview} for more details on LDA+DMFT. Let us now
turn to the spectroscopic data and our theoretical interpretation.

\section{Optical conductivity: Phase separation around the MIT}
\label{Sec:optcond}
The first work we will discuss, are measurements of
infrared optical conductivity \cite{Naturepap}. This experiment has a twofold goal: on the one hand, to clarify
the behavior of the 1.1\% Cr--doped compound around the metal to
insulator transition, and on the other hand, to perform an
experimental check of the pressure-doping equivalence. The motivation
of the former analysis is the following: In the past much
effort has been put into the understanding of the  transition between
the PM and the PI phase Fig.~\ref{V2O3_phase}. However, somehow less, or at
least less concrete, attention was paid to the local strain that
occurs in the lattice in the Cr--doped compounds
\cite{mcwhan69,frenkel06}, even though, for
(V$_{0.989}$Cr$_{0.011}$)$_2$O$_3$, the presence of a structural phase
separation, between the PM $\alpha$-- and PI $\beta$--phase, by the
Cr--doping has been  stated before \cite{mcwhan69,frenkel06,robinson75}.  Other experimental studies also support the idea that
the Cr-- atoms in (V$_{0.989}$Cr$_{0.011}$)$_2$O$_3$ could play the
role of $\beta$--phase ``condensation nuclei'':\\
Resistivity measurements, for example, show that the conducting phase of weakly
Cr--doped samples shows a bad metallic behavior, different from the
undoped compound \cite{kuwamoto80}.
Moreover, so called extended x--ray absorption fine--structure spectroscopy
(EXAFS) measurements showed that the presence of Cr contracts the Cr--V
bonds, inducing a concomitant elongation of V--V pair bonds
\cite{frenkel06}. Such ``long'' V--V pair bond is associated to the
$\beta$ PI phase \cite{mcwhan70}, as shown also by theoretical calculation using LDA+DMFT
\cite{held01B}. Therefore it may be hypothesized that, within an 
metallic matrix host, insulating--like ``islands'' are formed around
the Cr impurities \cite{cummings84,rice72}. On this basis, the PM--PI MIT has been suggested
to have also a percolative nature \cite{frenkel06}. 
Pressure--dependent transport studies by
Limelette \emph{et al.} \cite{limelette03} were also used to show that across the PI--PM
first order transition a large hysteresis occurs. This points to a
non--trivial role of the lattice and its distortions due to the Cr
doping, which has however been almost disregarded, or drastically
simplified when defining the phase diagram. The latter has been established
by means of resistivity data only and suggests the equivalence of doping and pressure.

The relation between such hysteresis and the above
mentioned coexistence of $\alpha$ and $\beta$--phases has not been 
clarified hitherto. It is these unknowns at which the more recent investigation \cite{Naturepap} aimed.

\begin{figure*}
  \begin{center}
  \includegraphics{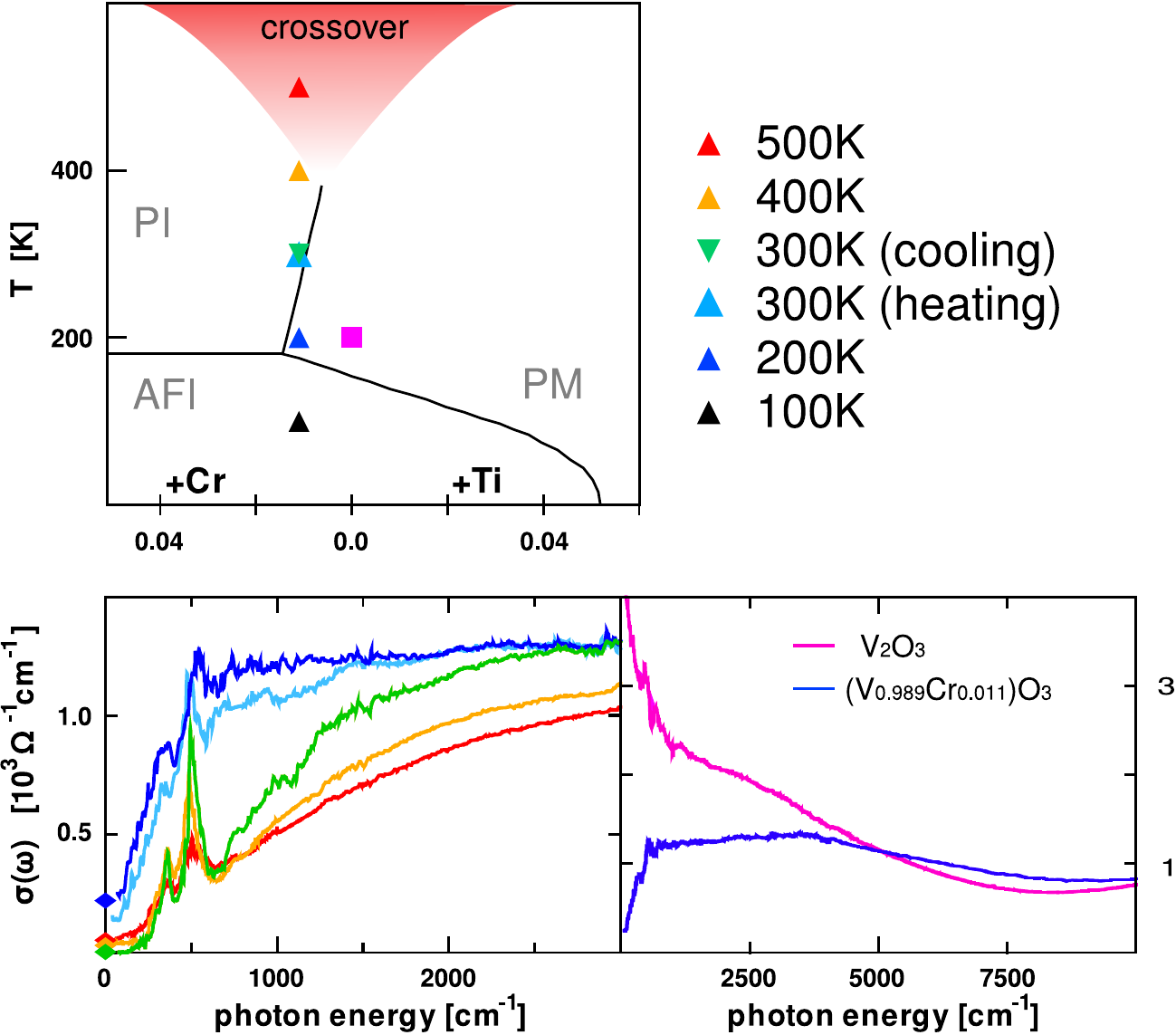}
  \end{center}
    \caption{(Color online)  IR optical conductivity for V$_2$O$_3$. Upper panel: phase
      diagram with marks at the positions where the spectra were
      taken. Lower panel left hand side: spectra for the 1.1\%
      Cr--doped (V$_{0.989}$Cr$_{0.011}$)$_2$O$_3$ sample at different
      temperatures ranging from 500K down to 200K. At 300K, hysteresis
      can be found and the spectra differ severely depending from
      which phase the point is approached. Lower panel right hand side:
      comparison between the (V$_{0.989}$Cr$_{0.011}$)$_2$O$_3$
      spectrum and the spectrum of undoped V$_2$O$_3$ at 220K
      (according to  \cite{Naturepap}).}
    \label{V2O3_optexp}
\end{figure*}

{\bf Experimental results }
The experimental results \cite{Naturepap} are presented  in Fig.~\ref{V2O3_optexp}: In the
upper panel the positions in the phase diagram where the
spectra were taken are marked. The spectra are plotted in the lower panel in the
corresponding color: on the left hand side several spectra for
different temperatures are shown together with their values for the DC
conductivity $\sigma_{\text{DC}}$ at $\omega\rightarrow 0$, whereas on the right hand side the T=200K
spectra for the undoped and the 1.1\% Cr--doped samples are compared.\\  
Let us start with the discussion of the temperature dependent
data. Shown in Fig.~\ref{V2O3_optexp} (lower panel left hand side) are
(V$_{0.989}$Cr$_{0.011}$)$_2$O$_3$--spectra in the temperature range
between 500K and 200K (the sharp peaks around 500cm$^{-1}$ correspond to
phonon resonances and are of no further interest for the present
discussion). Our starting point is in the cross--over region of the transition at
500K (red). Cooling down we obtain the spectrum at 400K (yellow) and
at 300K (green). At 300K, however, we are in the direct vicinity of
the transition line. Hence, we can, in fact, find a qualitatively
different (PM--phase) spectrum at 300K (light blue) if we approach the same point by heating up
from lower temperatures. That is, we observe hysteresis. The last spectrum we show in the plot was taken in the PM
phase at 200K (blue). The first three spectra, from 500K down to
300K (upon cooling) display the gapped shape which we expected as the
hallmark of the insulating nature of the PI $\beta$--phase. The spectra
show no Drude peak and, only at elevated temperatures, gain
minimal spectral wait at $\omega \rightarrow 0$. The remarkable, and
far from trivial, spectra are the ones in the PM phase at 300K (upon
heating, light blue) and at 200K (blue). We recall that already
resistivity measurements have shown a bad metallic behavior for the
Cr--doped sample as opposed to the undoped compound. Yet, how dramatic the difference to
the undoped sample really is, can be seen clearly in Fig.~\ref{V2O3_optexp}
(lower panel right hand side) where we show a direct comparison of the 
(V$_{0.989}$Cr$_{0.011}$)$_2$O$_3$--spectrum at 200K (blue) and the
spectrum of the undoped sample at the same temperature (pink): The spectrum of the undoped sample shows the behavior
that is expected from a metallic phase, including a well pronounced Drude peak.
In contrast,
the shape of the spectrum of the Cr--doped compound is
rather unexpected: It does not show a Drude
peak, but neither has it a gap like in the insulating regime as
the spectral weight around $\omega \rightarrow 0$ is non--negligible. This
fact is a clear support for a scenario of an inhomogeneous
(i.e. $\alpha$--$\beta$ mixed) metallic phase.
On the contrary, when comparing the behavior of
(V$_{0.989}$Cr$_{0.011}$)$_2$O$_3$ and (V$_{0.972}$Cr$_{0.028}$)$_2$O$_3$ within the PI
phase, only small differences appear (not shown here).

These new interesting experimental facts were motivation
enough  to revisit the compound again with the help of LDA+DMFT
in order to understand the features that are displayed more
fundamentally. Performing this analysis allows us to test the hypothesis
of the mixed $\alpha$--$\beta$ phase scenario \cite{Naturepap}.

{\bf LDA+DMFT analysis } 
The starting point for the theoretical LDA+DMFT analysis is the
downfolded NMTO Hamiltonian described in the previous section for the
$\alpha$-- and the $\beta$--phase respectively. In the
DMFT code, the Hirsch Fye quantum Monte Carlo method was employed. The calculations were carried
out at an inverse temperature of $\beta=20$ eV$^{-1}\approx 580$K and
with interaction parameters $U=4.0$ eV and $J=0.7$ eV.
After convergence of the DMFT self consistent loop the single particle
Green function on the imaginary time $\tau$ axis was analytically
continued by means of the Maximum Entropy Method \cite{MEM}. Next, the
local self energy was extracted on the real axis in order to calculate the optical
conductivity measured in the experiment. In the following we 
first discuss the direct DMFT results, i.e., spectral functions and local self energy
thereby also comparing them to the previous data from
Poteryaev \emph{et al.} \cite{poteryaev07}. Afterwards we present the
calculation of the optical conductivity.

{\bf  Interaction parameters for V$_2$O$_3$}
  From the technical perspective, we need to elaborate in more detail on the important issue of choosing the appropriate values
  for interaction parameters of a specific compound and the
  theoretical method that is employed.\\
  V$_2$O$_3$ presents a good example in that respect, since in the literature
  several different values for $U$ and $J$ can be found. The confusion
  about these parameters partly stems from the improvement in the
  estimates of their values over the time and partly from the
  differences in the numerical techniques.  A constrained LDA
  calculation (for the monoclinic antiferromagnetic phase) by
  I. Solovyev  \emph{et al.} \cite{solovyev96} yields the values of
  $U=2.8$ eV and $J=0.93$ eV -- parameters that later on were employed
  in some works \cite{ezhov99,held01,held01B}. Yet, constrained LDA gives
  unfortunately only very rough estimates of the values for U, which
  not only crucially depend on the electronic structures, but also on the basis set of the
  model at hand because it is highly sensitive to screening. For example, U has to be chosen much lower in the
  case of a LDA+U calculation (for V$_2$O$_3$ $U=2.8$ eV) in
  comparison to DMFT values ($U\approx 4$ eV) in order to overcome the
  deficiency of the static mean field  nature of LDA+U which
  overestimates ordering and gaps (see e.g. 
  \cite{sangiovanni06}). This leads to the parameter $U=4.0$ eV \cite{toschi09} following
  the philosophy of Held \emph{et al.} \cite{held01} and Poteryaev
  \emph{et al.}\cite{poteryaev07} that the
  value of $U$ should be consistent with the correct physics of
  V$_2$O$_3$, i.e., the MIT is reproduced within LDA+DMFT. Therefore,
  it is not surprising that this choice of $U$ agrees well with the previous
  LDA+DMFT studies.  The
  result of our analysis clearly demonstrates that the appropriate $U$
  for the LDA+DMFT calculation should be chosen in the range $4.0$ eV
  $<U<4.2$ eV\cite{toschi09}, as we did in the present calculations. Considerably
  smaller and larger values of U would either lead to the disappearance or a
  huge overestimation of the spectral, and as to be seen also optical,
  gap in the PI phase.

\begin{figure*}
  \begin{center}
  \includegraphics{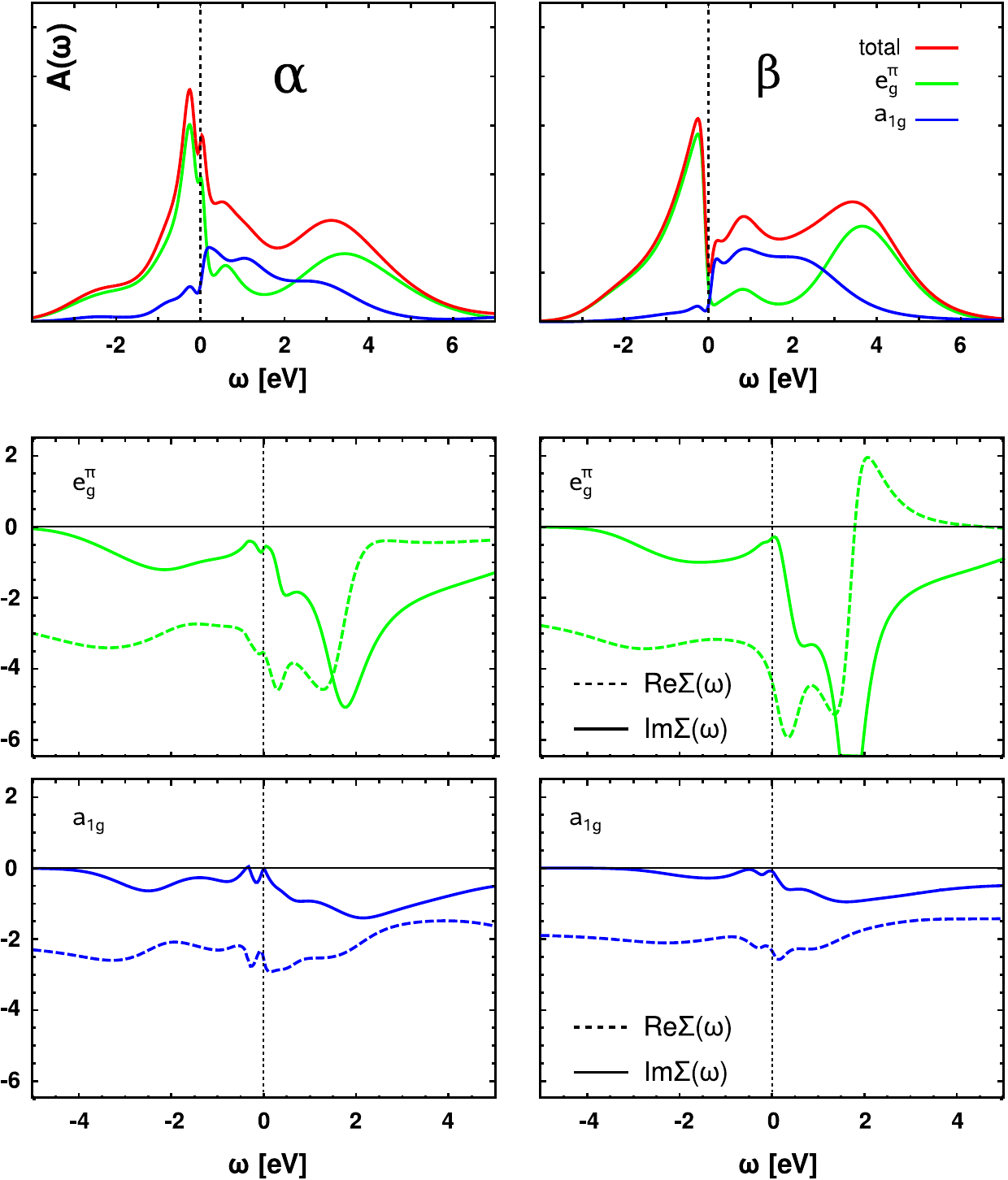}
  \end{center}
    \caption{(Color online) LDA+DMFT results for (V$_{0.989}$Cr$_{0.011}$)$_2$O$_3$ in
      the $\alpha$ phase (left hand side) and the $\beta$ phase (right
      hand side) for $U=4.0$ eV, $J=0.7$ eV and inverse temperature $\beta=30$eV$^{-1}$. In the upper panels we show the LDA+DMFT spectral
      functions resolved in orbital labels and coded by their color
      (see legend). The spectra show metallic behavior (coherent
      excitations around the Fermi energy) for the
      $\alpha$ phase and gapped insulating behavior for the $\beta$
      phase. In the lower panels we show the self energies
      for the $e_g^\pi$ (middle) and $a_{1g}$ (bottom) states.}
    \label{V2O3_awsfspec}
\end{figure*}

{\bf LDA+DMFT results for (V$_{0.989}$Cr$_{0.011}$)$_2$O$_3$ }
In Fig.~\ref{V2O3_awsfspec}, we report  orbital--resolved spectral
function (upper panel) as well as the corresponding self energies (lower
panels). In the plots we set the Fermi energy to
$\varepsilon_{\text{F}}=0$ and plot the sum of the two degenerate
$e_g^\pi$ spectra in green, the $a_{1g}$ spectrum in blue, and the
total spectrum, i.e., the sum over all, in red color. We
summarize the quantities for the $\alpha$-- and the $\beta$--phase on
the left hand and right hand side of the panels respectively.
Overall our results agree with the results of the previous LDA+DMFT
analysis by Poteryaev \emph{et al.} \cite{poteryaev07}, although we
performed the calculations at slightly lower $U=4.0$ eV values (in
\cite{poteryaev07} $U=4.2$ eV)\footnote{The reason for our choice is
  the sensitive dependence of the optical gap on this parameter.}. The
self energies, both real and imaginary parts, display a strongly
orbital dependent character. The real part acts like an orbital
dependent renormalization of the chemical potential or, in other
words, as it is called in Refs.\ \cite{keller04,poteryaev07} as an ``effective crystal
field''
 whereas the imaginary part is a
measure of lifetime/coherency of the excitations in the associated
band. However, the self energy depends also on the filling of the
respective orbitals and in a hybridized system like the $t_{2g}$
states of V$_2$O$_3$ it is a very involved quantity: Although the
self energy is diagonal we see from equation \eqref{V2O3_gloc} that its
connection to the Green function, and hence the spectrum, involves an
inversion so that the orbitally resolved information is, in a way,
convoluted.\\
The spectral functions for the $\alpha$ and $\beta$--phase
are quite similar, except for the strongly renormalized coherent
quasiparticle excitations of the correlated metallic $\alpha$--phase
around the Fermi energy. Of course, the differences are expected to be
sharpened up at lower temperatures. From the orbital--resolved
spectra we can obtain valuable insight. 
Let us have a closer look at the incoherent part of the spectrum,
i.e., the Hubbard bands. The basic features can be understood as
follows: As it was discussed in previous works (e.g. \cite{held01}) and also will be
confirmed later by the XAS study \cite{rodolakis10} the predominant local configuration
on the V atoms has two spin--aligned electrons in the $e_g^\pi$
orbitals, i.e., a $\ket{e_g^\pi e_g^\pi}$ spin--1 configuration, with
some admixture of $\ket{a_{1g}e_g^\pi}$ spin--1
configurations. \label{discussJ} For a simple picture let us first
consider the lower Hubbard band (LHB), that is, the electron removal part
of the spectrum. We recall the relevant onsite interaction parameters
to be the
intra--orbital interaction $U$, the inter--orbital interaction $V$,
and the spin--coupling constant $J$. Furthermore, in cubic (or close to cubic)
symmetry the relation $V=U-2J$ holds.
Starting either from the $\ket{e_g^\pi e_g^\pi}$ or the
$\ket{a_{1g}e_g^\pi}$ configuration, the removal of an electron will
result in an energy gain of $V-J$   ($\approx 1.9$ eV in our case) which is in
agreement with the position of the LHB. The only structure,
i.e. splitting,  which occurs is the crystal field potential
differences of the $e_g^\pi$ and $a_{1g}$ states. This energy scale,
however, is below the resolution of our spectra at high $|\omega|$. For the upper Hubbard
band, i.e., the electron addition part, the situation turns out to be a
little bit different. The additional electron  can either populate an
$e_g^\pi$ or an $a_{1g}$ state. Then the
process $\ket{e_g^\pi e_g^\pi}\rightarrow\ket{e_g^\pi e_g^\pi
  e_g^\pi}$ or $\ket{a_{1g} e_g^\pi}\rightarrow\ket{a_{1g} a_{1g}
  e_g^\pi}$ will cost an energy of $U+V$. The additional electron
interacts via $U$ with one of the other two electrons, and via $V$
with the other one. Transitions
$\ket{e_g^\pi e_g^\pi}\rightarrow\ket{e_g^\pi e_g^\pi 
  a_{1g}}$ or  $\ket{a_{1g} e_g^\pi}\rightarrow\ket{e_g^\pi e_g^\pi
  a_{1g}}$ only cost $2V$ or $2V-2J$ depending on the respective
spin alignment. Consequently, the UHB is split into two main
features which we can find around $1$ eV and $4-5$ eV. We conclude that i) the
split of the UHB depends apparently strongly on the choice of $J$ and
ii) this splitting is responsible for the small width of the gap
compared to the interaction parameters\footnote{This observation explains also why the attempt to handle the gap (actually the optical
gap) with a one band Hubbard model \cite{rozenberg95} led to
unphysically small values for the interaction parameter.}.

\begin{figure*}
  \begin{center}
  \includegraphics{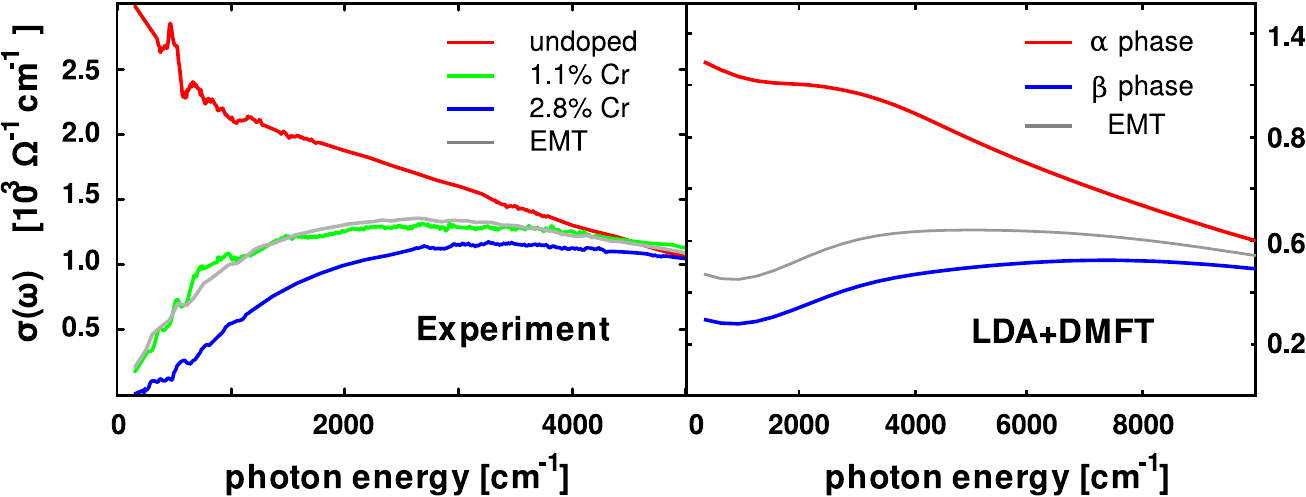}
  \end{center}
    \caption{(Color online) Phase mixing: Comparison of experimentally observed (left hand side)
      and LDA+DMFT calculated (right hand side) optical
      conductivity. In red we plot the spectrum for the $\alpha$ phase,
      for which we take the experimental spectrum of the undoped
      sample \cite{Naturepap}. The $\beta$ phase, for which we take the experimental
      spectrum of the 2.8\% Cr doped sample is shown in blue.
      The measured spectrum of the 1.1\% Cr doped sample (green) can
      be fitted by a mixture of $\alpha$ and $\beta$ phase spectra
      (gray) within the \emph{effective medium theory}   \cite{Naturepap}.}
    \label{V2O3_optcond}
\end{figure*}

Let us, finally, turn to the optical conductivity. The LDA+DMFT calculation of
the optical conductivity has been performed within the Peierls approximation
and neglecting vertex corrections
along the line of 
\cite{tomczakthesis}.\footnote{It should be remarked, that the non--monoatomic
basis of the crystal leads to corrections in the calculation of the
Fermi velocities even at the level of the Peierls approximation as it is discussed
by Tomczak \cite{tomczakthesis} and Tomczak and Biermann \cite{tomczak09}. For V$_2$O$_3$, however, such corrections only concern the optical conductivity along the $c$ axis not in the xy-plane.}\\
In Fig.~\ref{V2O3_optcond} we show a comparison of experimental data
(left hand side) and LDA+DMFT data (right hand side).
We calculated the optical conductivity also for both $\alpha$ and $\beta$--phase. The LDA+DMFT
optical conductivity of the $\beta$--phase (right hand side: blue)
shows a gapped behavior, as it is expected for the PI phase. The
fact that it does not extrapolate to zero at energies lower than
$1000$cm$^{-1}$ is due to the temperature of $\beta^{-1}=0.05 {\rm
  eV}\approx 500$ K assumed for the DMFT(QMC) calculations. Further, when we
compare it to the experimental data of the 2.8\% Cr--doped sample,
deep in the PI phase, we see that the calculated gap of \cite{Naturepap}
$\sigma_{\beta}$ is a little bit too large. The reason for this overestimation is an
extreme sensitivity of the calculation on the choice of $U$ and $J$ as
it was already mentioned before\footnote{A slightly larger $U$, like it
was used, e.g., in \cite{poteryaev07} would result in an even larger
gap.}. The calculated $\alpha$--phase optical conductivity (right hand
side: red) shows an overall good agreement with the experimental data
taken for the undoped compound (left hand side: red). At
$\omega\rightarrow 0$ we can distinguish the typical Drude peak contribution of
the PM phase.\\ 
The most interesting spectrum, however, corresponds to the experimental data taken
for the 1.1\%Cr--doped sample at 200K (left hand side: green): As
mentioned above this spectrum is strange in its shape (with neither
Drude peak nor gap) and belongs to a state that is, according to the
resistivity measurements, a bad metal. The discrepancy between the idea
that the PM phase can be seen as a uniform metallic phase and the
experimental evidence is further enhanced by our LDA+DMFT
calculations. Specifically, as the lattice parameters practically do not
change within the $\alpha$--phase, the LDA+DMFT spectrum of the
1.1\%Cr--doped sample at 200K and of the undoped compound are
basically indistinguishable. Hence, our calculations support the 
hypothesis of an $\alpha$--$\beta$ phase mixture in the bad metal
region. To test this hypothesis further we resort to a
semi--empirical formula of the \emph{effective medium theory}
(EMT) \cite{cummings84,carr85} which provides a simple way of
approximating the dielectric constants
$\bar{\varepsilon}(\omega)=(1+4  \pi i \sigma(\omega)/\omega)$ for mixtures of insulating
and metallic phases. Within the EMT the effective constant
$\bar{\varepsilon}(\omega)_{\text{eff.}}$ satisfies the condition\\
\begin{equation}\label{EMT}
f\frac{\bar{\varepsilon}_{\text{met.}}(\omega)-\bar{\varepsilon}_{\text{eff.}}(\omega)}{\bar{\varepsilon}_{\text{met.}}(\omega)+\frac{1-q}{q}\bar{\varepsilon}_{\text{eff.}}(\omega)}+(1-f)\frac{\bar{\varepsilon}_{\text{ins.}}(\omega)-\bar{\varepsilon}_{\text{eff.}}(\omega)}{\bar{\varepsilon}_{\text{ins.}}(\omega)+\frac{1-q}{q}\bar{\varepsilon}_{\text{eff.}}(\omega)}=0
\end{equation}

Where $f$ and $q$ are free fitting parameters which are
phenomenologically related to the size and relative densities of the
``islands'' of the two constituent phases. For further information
about this approach we refer the reader to \cite{basov07}. Now we take the
optical conductivity spectra measured in the undoped sample and the 2.8\%
Cr--doped sample as the $\alpha$-- and $\beta$--phase spectra
respectively and use Eq.~\eqref{EMT} to fit the experimental spectra of the 1.1\%
Cr--doped sample. For the values $f=0.42$ and $q=0.35$ an
excellent agreement can be found which is plotted in Fig.~\ref{V2O3_optcond}
(left hand side: compare green and gray).
From a theoretical point of view, it is even possible to directly use
the LDA+DMFT spectra for the $\alpha$-- and $\beta$--phase of
(V$_{0.989}$Cr$_{0.011}$)$_2$O$_3$ as an input for the EMT. Also in
this case, with the same values of $f$ and $q$ we obtain a satisfying
agreement with the experimental data.\\
To sum up, the experimental measurements of the
optical conductivity together with the theoretical interpretation by
means of LDA+DMFT strongly support the scenario of a mixed phase state
for the 1.1\% Cr--doped compound at 200K. It will be seen in 
Section \ref{Sec:xray} that the complementary x--ray absorption spectroscopy also speaks for this
scenario.

\begin{figure*}
  \begin{center}
  \includegraphics{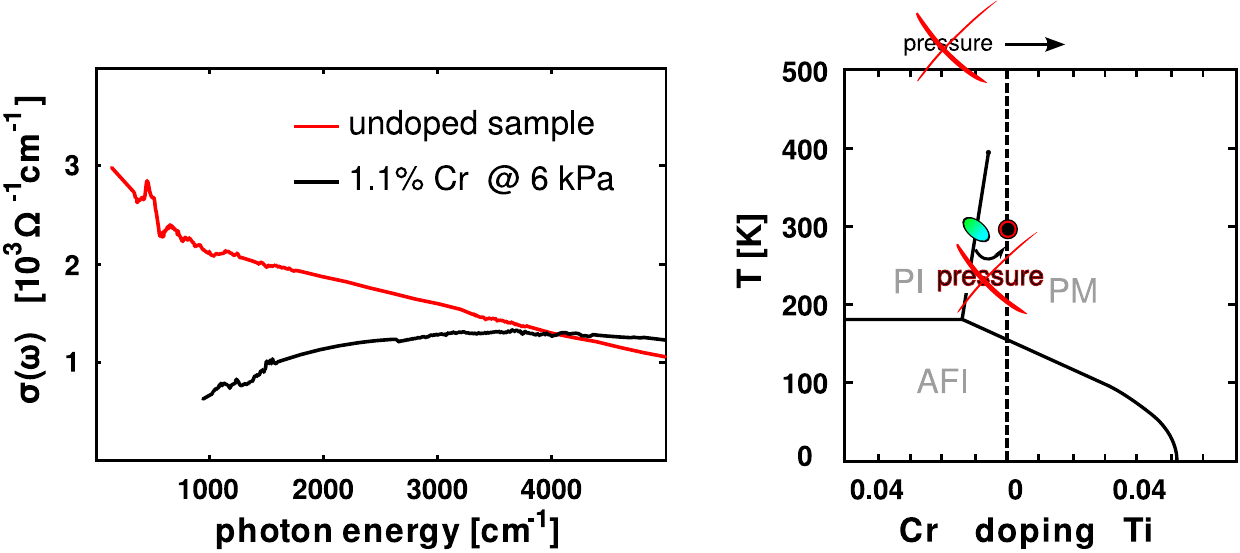}
  \end{center}
    \caption{(Color online) Left: Experimental measurements of the optical conductivity for the
    undoped sample (red) and the 1.1\% Cr doped sample at 6 kbar \cite{Naturepap}. For
    this pressure the ``common wisdom'' of a pressure-doping equivalence would predict  a fully recovered PM
    phase equivalent to the undoped compound PM phase. Right: The
    experimental data clearly proofs this assumption to be incorrect
    and suggests to abandon the concept of pressure equals inverse doping. } 
    \label{V2O3_purevspressure}
\end{figure*}

The last part of our discussion about the optical conductivity is
devoted to the data of the 1.1\% Cr doped sample under pressure far in
the metallic region at 6kbar. As it was stated in the
beginning, and motivated by the results we already discussed, the
second question we want to address is whether the doping with Cr can
really be ``reversed'' by applying an external pressure. In short: Can
pressure really be drawn on the same axis in the phase diagram as the
doping? Experimental results from optical spectroscopy give a clear negative answer to that
question. In Fig.~\ref{V2O3_purevspressure} we report on the left hand
side the comparison of the experimentally measured spectra for the
undoped and the 1.1\% Cr--doped sample \cite{Naturepap}. The spectra are,
evidently, not even qualitatively similar. This indicates the existence of
different PM states obtained by either tuning temperature/doping or applying
pressure.\\
It remains, however, to formulate and quantify this difference in a
rigorous manner. This will be the subject of the section \ref{Sec:xray}, in which we
discuss the hard x--ray absorption spectra on the vanadium K--edge.

\section{Photoemission microscopy}
\label{Sec:photomicro}
\begin{figure*}
  \begin{center}
  \includegraphics[width=16cm]{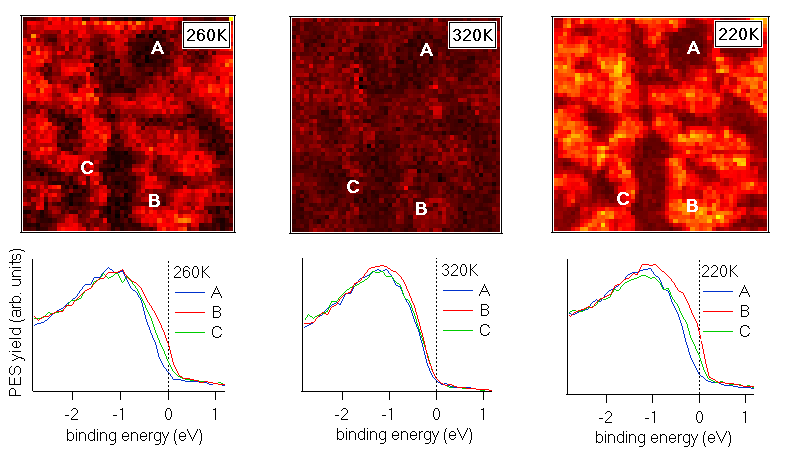}
  \end{center}
    \caption{(Color online) Representative intensity ratio images and corresponding
      photoemission spectra at different temperatures over a 50 $\mu$m
      by 50 $\mu$m area obtained by scanning photoemission 
      microscopy using photons at $27$ eV \cite{Naturepap}. Inhomogeneous properties
      are found within the PM phase at 220 or  260 K, where
      metallic (bright) and insulating (dark) domains coexist. A
      homogeneous insulating state is instead obtained in the PI phase
      at $320$ K. After a whole thermal cycle the structure of the
      inhomogeneous distribution is recovered, indicating the presence
      of stable ``condensation nuclei''.}
    \label{V2O3_microscopy}
\end{figure*}

Recently, also spatially resolved photoemission microscopy
data was obtained for the 1.1\% Cr--doped compound \cite{Naturepap}. 
This experimental technique makes it possible to obtain the detailed physical
information of photoelectron spectroscopy with a lateral resolution
of the order of 100 nm \cite{Gregoratti}, and has already proven its capability of 
spatially resolving inhomogeneous electronic structures \cite{microbands} that would otherwise give an average signal with macroscopic probes \cite{macrobands}. 
Applied to the 1.1\% Cr--doped compound, this method provides
a spectacular confirmation of our interpretation based on a mixed
phase state in the paramagnetic metallic phase: in
Fig.~\ref{V2O3_microscopy} we show the microscopic images together
with the PES for the labeled positions. At 260K one clearly observes a
mixture of areas with coherent excitations (with finite spectral weight at
the Fermi energy) and insulating regions (gapped spectra) as we would
expect in our scenario. Upon heating, the system
becomes completely insulating (compare the image at 320K). Note,
after cooling down to 220K again the same ``map'' as before is
recovered.
Note,  differences in Fig.~\ref{V2O3_microscopy} between 220 and 260K are due to
the different temperatures, not due to the hysteresis loop,
see the additional data in \cite{Naturepap}.
This memory effect becomes even more evident in a similar study
on the phase transition between the paramagnetic metallic and the antiferromagnetic 
insulating phase \cite{MansartAPL}. This suggests a correlation between 
the position of the insulating regions and the nucleating action of structural defects
in the material, which tend to guide the natural tencency of this system 
towards phase separation. The structural defects may well be related to 
the lattice strain caused by the presence of Cr--impurities in the material, 
even though this conjecture has to be further clarified both experimentally and  theoretically. \\

\section{X--ray absorption on the V--K--edge: Pressure vs. doping}
\label{Sec:xray}

Among the different experimental methods recently employed to study 
the electronic properties  of the Mott transition in Cr--doped 
V$_2$O$_3$~\cite{limelette03,mo03,mo06,rodolakis10,rodolakis09B}, X-ray 
absorption spectroscopy (XAS) has played a crucial role. For instance, it was the 
detailed investigation of the V $L_{2,3}$ absorption 
edges~\cite{park00} that demonstrated the necessity of abandoning 
the simple one band, $S=1/2$, model to obtain a realistic description 
of the changes in the electronic structure at the phase
transition. Further, Park \emph{et al.} obtained valuable quantitative information
about the vanadium ground state for different amounts of doping and
temperatures \cite{park00} and formulated it as a linear combination of the
$\ket{e_g^\pi e_g^\pi}$ and the $\ket{a_{1g} e_g^\pi}$ states which
were mentioned earlier. This kind of tool would be perfect to also
clarify the question which remains from the discussion of the previous
section: What is the character of the metallic ground state of the
Cr--doped sample under pressure? However, unfortunately the V
$L_{2,3}$ absorption falls in the region of soft x--ray radiation, and
thus, due to the specific absorption characteristic of the diamond
anvil cell used for the pressure measurements, it cannot be employed in our
case.
But fortunately XAS can also be performed at the V $K$--edge in the
hard x--ray range, i.e., in a spectral region without particular
absorption of the diamond anvil cell. In this case, the pre--edge will carry most of the
physical information we are interested  in, as it is predominantly due
to $1s\rightarrow 3d$ transitions. The excitations in this pre--edge
region are influenced also by the core hole and should be considered to be 
of an excitonic nature. Beside the possibility of measuring the V
$K$-edge under pressure condition we obtain also a more straightforward
interpretation. Namely, due to the simple spherical symmetry of the
$s$-core hole, the  multiplet structure reveals a more direct view on the
$d$-states.\\
Motivated by the above considerations, we used in Ref.\  \cite{rodolakis10,rodolakis11} V $K$-edge XAS to explore extensively the MIT in V$_2$O$_3$ by 
changing temperature, doping and applying an external pressure. The 
onsets of the $K$-edges were analyzed by a novel computational scheme combining 
the LDA+DMFT method with configuration interaction (CI) full multiplet
ligand field
calculations  to interpret subtle differences at the PM--PI 
transition. 

Such CI, or as they are frequently called 'cluster calculations' became within the last 25 years
a popular parameter based method for fitting experimental data in order to extract information
on charge, spin, and orbital degrees of freedom particularly in correlated transition metal oxides (see e.g. Refs. \cite{tanaka94,groot94,thole97,haverkortthesis}).
In this work, however, we do not fit to experiment, but instead use parameters derived from an \emph{ab initio} LDA+DMFT scheme
in order to calculate the experimentally measured x-ray spectrum (for further, more technical details see Ref.~\cite{hansmann12a}).

This analysis allowed us to: (i) observe in detail the changes in
This allowed us to: (i) observe in detail the changes in 
the electronic excitations, providing also a direct estimate of the
Hund's coupling $J$ (recall the discussion of the LDA+DMFT spectral
functions in the previous chapter) (ii) analyze the physical
properties of the PI and PM phase on both sides of the MIT, leading to
the main result of our work: (iii) understand the difference between
P, T or doping-induced transitions. This difference is mainly related
to the occupancy of the $a_{1g}$ orbitals, suggesting the existence of
a new ``pressure'' path between PI and PM in the phase diagram, which is distinctive from the "doping'' path. The
XAS is in that respect complementary to the optical conductivity
measurements.  The optical conductivity is
connected to somewhat non--local excitations. Therefore it was a great
tool to confirm the mixed phase scenario. In contrast the XAS, or more
specifically the excitonic features of XAS offers us information about the $d$ occupations  from
a completely localized perspective which is needed in order to formulate
the ground state in the language of localized Wannier orbitals. This information in turn
                     could not be extracted from the optical
conductivity.\\
For the experiments  high quality samples of
(V$_{1-x}$Cr$_x$)$_2$O$_3$ with various doping in the PM ($x=0$) and 
PI phases ($x=0.011$ and $0.028$) at ambient conditions were used in \cite{rodolakis10}. The MIT was  
also crossed for the $0.011$ doping by changing temperature and for the 
$0.028$ doping by pressure. To obtain the best resolution, the XAS 
spectra were acquired in the so-called partial fluorescence yield 
(PFY) mode~\cite{groot01}, monitoring the intensity of 
the V-K$\alpha$ 
($2p\rightarrow 1s$) line as the incident energy is swept across the 
absorption edge.
Further experimental details can be found in
\cite{rodolakis10,rodolakis11}.

\begin{figure*}
  \begin{center}
  \includegraphics{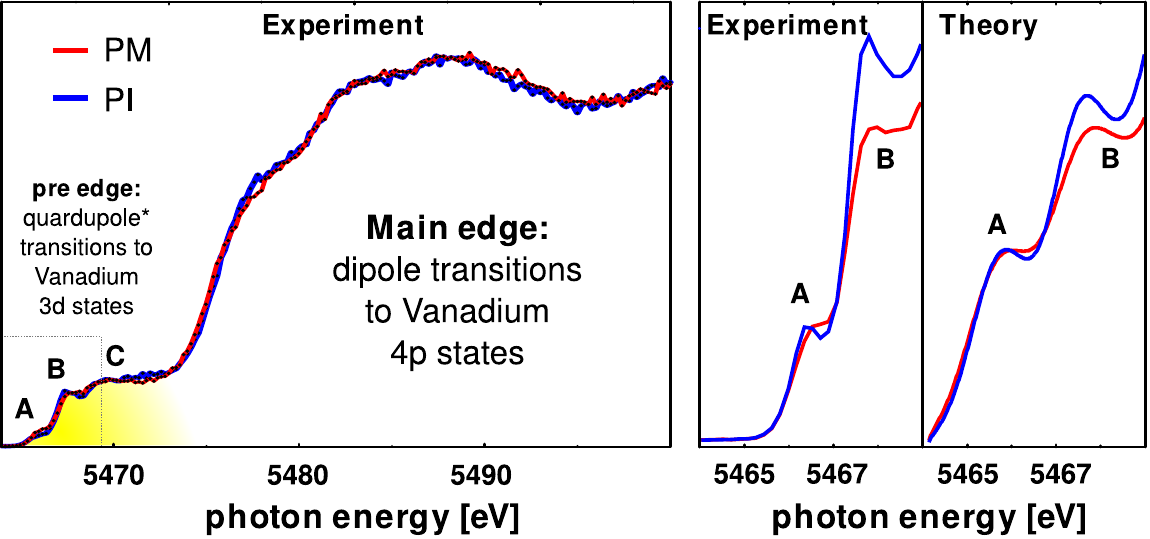}
  \end{center}
\caption{(Color online) Vanadium $K$-edge x-ray absorption spectra in 
  (V$_{1-x}$Cr$_x$)$_2$O$_3$ for a powder sample with $x=0.011$ measured as a 
  function of temperature (T) in the PM (200 K, red line) and PI 
  (300 K, blue line) phase by partial fluorescence yield XAS. In the
  region above 5475eV the main edge starts; here dipole transitions from the core electron to
  the vanadium 4p states give the main contribution. Below 5470eV we
  find the pre--edge -- where, besides others, the transitions to the
  vanadium 3d states are located. From these we extract the information about the
  ground state of the system. On the right hand side we show a zoom of
  the pre--edge region and compare the structure to theoretical
  full multiplet CI spectra. As explained in the text, these
  transitions would be pure quadrupole if it would not be for the
  inversion symmetry breaking on the vanadium site that makes the
  transitions ``slightly dipole allowed''. (reprinted with permission from  \cite{rodolakis10}
 Copyrigth (2010) by the American Physical Society))}
\label{V2O3_powderxasmainandpre}  
\end{figure*}  

{\bf Powder data and isotropic calculations }
The T--dependent absorption spectra are displayed in 
Fig.~\ref{V2O3_powderxasmainandpre} left hand side for both PM (200 K) and PI (300 K) phases for the 
$x=0.011$ powder sample. The spectra have been normalized to an edge 
jump of unity. We will focus on the pre--edge region, where 
information about the V $d$-states can be extracted as it is indicated
in the plot. It can be decomposed into three spectral features (A,B,C)
which all vary in intensity as the system is driven through the MIT
whereas C is considerably broader then A and B. Notice that no 
feature is observed below peak A contrary to the early results of 
Ref.~\cite{bianconi78} but in agreement with the more recent data of 
Ref.~\cite{goulon00}. Within a simplified atomic like picture, one 
could directly relate the intensity of features A,B and C to the 
unoccupied states: The V--$t_{2g}^2$ states are split into one 
$a_{1g}$ and two $e_g^\pi$ states under trigonal distortion of the V sites \cite{tanusri09}
 as shown in Fig.~\ref{V2O3_lvl}. Starting from a V--$t_{2g}^2$, $S=1$ configuration, 
one can either add an electron to the $t_{2g}$ subshell yielding peaks A and B, or add an electron to the $e_g^\sigma$ 
sub--shell which gives rise to the broader peak C. In this picture, 
Hund's rule exchange splits peaks A and B into a quartet ($S=3/2$) and
doublet ($S=1/2$) states.\\
This point of view is, however, oversimplified as the V $d$ electrons are 
{\it strongly correlated} and, in the pre--edge region, the spectra are
still largely influenced by the $1s$ core hole potential. Keeping that
in mind, we have simulated in  \cite{rodolakis10,hansmann12a} the pre--edge by combining CI with LDA+DMFT
calculations for which the one particle part (LDA) input corresponds
to the level diagram in Fig.~\ref{V2O3_lvl}. We concentrate our analysis to peaks A and B, since peak C relates mainly 
to the unoccupied $e_g^\sigma$ orbitals. These hybridize much stronger 
with the ligands and thus lack direct information on the 
Mott transition; peak C may also be related to 
non--local excitations (not included here)~\cite{gougoussis09} which 
sensitively depend on the metal--ligand distance.
Let us also note that the V sites in V$_2$O$_3$ are non 
centro--symmetric which leads to an on--site mixing of V-$3d$ and
V-$4p$-orbitals and interference between dipole and quadrupole
transitions~\cite{elfimov01}. This interference has been included in our scheme,
see \cite{hansmann12a} for  details, together with the linear
dichroism measurements.\\
The CI calculations  \cite{rodolakis10,hansmann12a} confirm  that for the ground state the occupancy ratio between the 
($e_{g}^{\pi}$,$a_{1g}$) and ($e_{g}^{\pi}$,$e_{g}^{\pi}$) states is smaller in 
the PI than in the PM 
phase~\cite{keller04,park00}: The isotropic 
CI--based calculated XAS spectra in the pre-edge region reported in 
Fig.~\ref{V2O3_powderxasmainandpre} right hand side agree well with the experimental data for both  
the energy splitting of features A and B and the ratio of their 
spectral weight (SW) which increases in the PM phase.

\begin{figure}
  \begin{center}
    \includegraphics{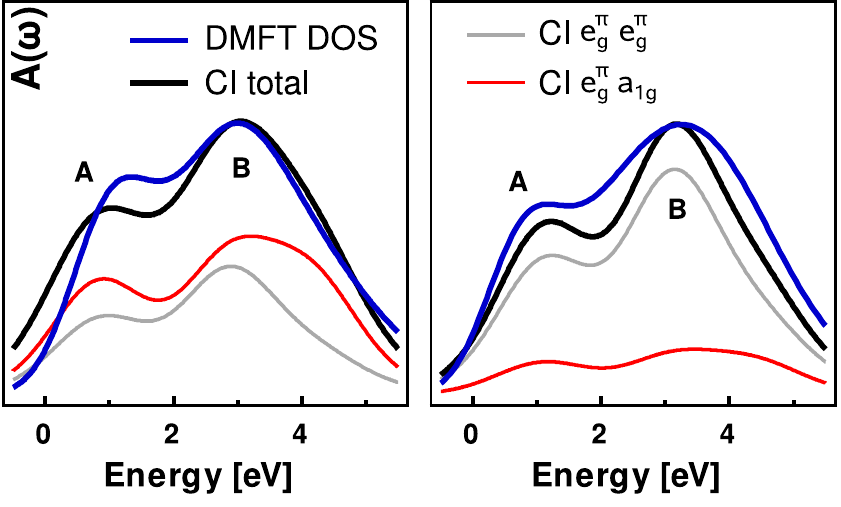}
  \end{center}
  \caption{(Color online) Spectral function $A(\omega)$ comparing 
    ``incoherent'' LDA+DMFT part (blue line) and the CI calculation (black line) in the 
  pre-edge region; left panel: PM, right panel: PI phase; $\varepsilon_{\rm F}=0$ is the Fermi energy. Note the similarity in 
  the main spectral features when crossing the MIT. Also shown are the 
  different  contributions of the CI spectrum labeled accordingly to 
  their initial state: the contribution  of the 
  ($e_{g}^{\pi}$,$a_{1g}$)~$\rightarrow$~($e_{g}^{\pi}$,$e_{g}^{\pi}$,$a_{1g}$) 
  transitions  (red line) to the peak A(B) is approximately $60\%(55\%)$ in the PM 
  phase and $20\% (15\%)$ in the PI phase  (reprinted with permission from \cite{rodolakis10}
 Copyrigth (2010) by the American Physical Society).} \label{V2O3_dmftvsCI}
\end{figure}

Considerable insight can be gained by comparing CI and LDA+DMFT 
calculations. Our LDA+DMFT calculations \cite{rodolakis10}, performed using 
the same NMTO Hamiltonian 
with the 1.1\% Cr--doped V$_2$O$_3$ and Hirsch--Fye Quantum Monte Carlo as impurity 
solver, confirm the above mentioned tendency. Specifically we obtain
a mixing  of $50$:$50$ and $35$:$65$ for the
($e_{g}^{\pi}$,$a_{1g}$):($e_{g}^{\pi}$,$e_{g}^{\pi}$) occupation in
the PM and PI phases respectively.
Remarkably the simple structure of the core hole 
potential in the $K$-edge spectroscopy ($L=0$ i.e. spherical potential) allows us to associate the 
pre--edge spectrum with the $k-$integrated spectral function above the 
Fermi energy calculated by LDA+DMFT in which we do not take into
account the core hole effects. The electron--addition part of the
spectral function shows three main features in PM phase: a 
coherent excitation at the Fermi level and a much broader double 
peak associated to the incoherent electronic excitations, i.e., the
upper Hubbard band (UHB), almost identically to 
the undoped compound. In the PI phase obviously, only the latter 
survives. Comparison with the experimental spectra clearly 
shows that the pre--edge features have to be related to the 
``incoherent'' part of the spectral function only. The physical reason is that the 
core hole potential localizes the electrons destroying 
the (already strongly renormalized) coherent quasiparticle excitations and making the XAS spectrum 
atomic--like. The ``incoherent'' LDA+DMFT, CI and experimental 
spectra shown in Figs.~\ref{V2O3_dmftvsCI} and \ref{V2O3_powderxasmainandpre}, respectively, 
agree in many aspects, especially as for the splitting of the 
first two peaks by $\approx$2.0 eV ($\approx$1.8 eV in experiment) which 
originates in LDA+DMFT from the Hund's exchange $J$ in the Kanamori 
Hamiltonian (see discussion of the LDA+DMFT spectral functions in Section
\ref{discussJ}). This further validates the choice of $J=0.7$ eV used in our 
calculations in contrast to larger values assumed in previous  
studies~\cite{keller04,laad06}, and also clarifies 
the mismatch between XAS and LDA+DMFT spectra reported in the undoped 
V$_2$O$_3$ compound~\cite{keller04} where incoherent excitonic
features were identified by coherent quasiparticle excitations. Moreover, the ratio between 
A and B peak displays the same trend in the PM-PI 
transition as the CI (or experimental) data. The quantitative 
difference between the two calculations is attributed to the lack of 
matrix elements in LDA+DMFT.\\
The intensity ratio of the first two incoherent excitations peaks A
and B (associated to the quartet and doublet states in the oversimplified picture) 
thus appears as the key spectral parameter to understand the
differences between PM and PI. Even in a powder sample, this ratio is
still sensitive to the $a_{1g}$ orbital occupation of the initial
state. Indeed, due to the
trigonal distortion a considerable spectral weight transfer from the
peak B to higher energies (corresponding to final states with two
$a_{1g}$ electrons) can take
place for the ($e_g^{\pi},a_{1g}$)  but not for the
($e_g^{\pi},e_g^{\pi}$) initial state. Therefore, the $K$ pre--edge XAS
can serve as a direct probe of the $a_{1g}$ orbital occupation in the 
ground-state. As a rule of thumb, the larger the ratio between the SW 
of A and B, the larger the $a_{1g}$ orbital occupation.

\begin{figure}
\sidecaption
    \includegraphics{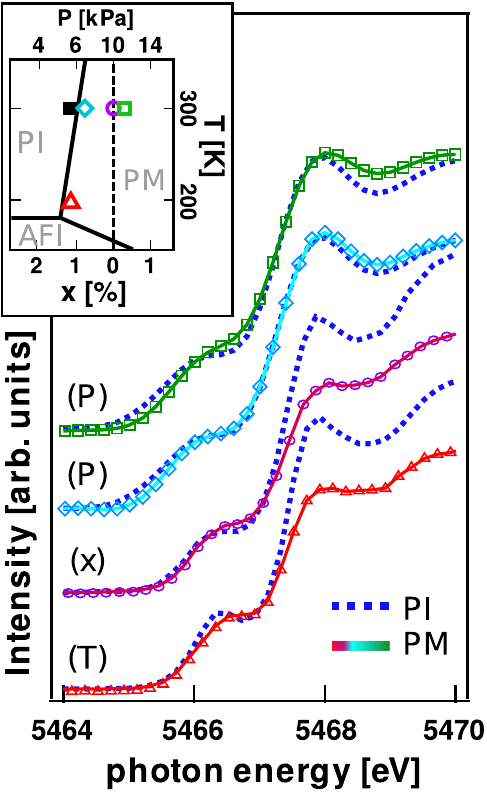}
  \caption{(Color online) V--K edge XAS spectra for powder samples of
      (V$_{1-x}$Cr$_x$)$_2$O$_3$, starting from the top, as a function
      of pressure (P) ($\square$) [$x=0.028$; 
      5 and 11 kbar (topmost set of curves); 5 and 7 kbar (second
      highest set of curves)],
      temperature ($\triangle$) [$x=0.011$; 200,300 K (second lowest
      set of curves)] (T), and doping 
      ($\circ$)  [$x=0$, 0.011 (bottom curves)] ($x$) (cf.\ points in the phase diagram; the pressure 
      scale refers to the $x=0.028$ doping). The spectral differences 
      demonstrate the nonequivalence between $P$ and temperature-doping. The $x$-$T$ 
      equivalence is confirmed by the photoemission spectra~\cite{rodolakis09}. \label{V2O3_pressurexas}}

\end{figure}

{\bf Under pressure }
After we established an interpretation scheme of the vanadium
XAS $K$--edge which allows us to use it as a ground state occupation probe it is
time to come back to the original task of inquiring the metallic phase
of Cr--doped V$_2$O$_3$.
Fig.~\ref{V2O3_pressurexas} shows the XAS 
powder spectra of the pressure--induced MIT with the corresponding 
spectra for the temperature-- and doping--driven transition (the markers in the 
phase diagram, Fig.~\ref{V2O3_pressurexas}). We remark at this point
that the spectra taken under pressure display a relative shift
between main--edge (not shown) and pre--edge, which is in any case
irrelevant for our discussion of the ground state for which we only
need the intrinsic structure of the pre--edge. Hence, this shift is
compensated for the pressure spectra in Fig.~\ref{V2O3_pressurexas}.
Fig.~\ref{V2O3_pressurexas} clearly evidences that (besides the shift) 
contrary to the doping- or T-driven transition, very small 
changes in spectral shapes and weights are observed in the pressure driven MIT.  
In the light of the arguments discussed 
above, our finding proves that the metallic state reached by applying 
pressure is characterized by a much lower occupation of the $a_{1g}$ 
orbitals compared to the metallic state reached just by changing 
temperature or doping. Importantly, the spectra measured 
through the doping induced MIT are identical within the experimental 
uncertainty to those measured through the temperature driven
transition. The 
temperature--doping equivalence is confirmed by photoemission data
\cite{rodolakis09} and is consistent with the very similar lattice
parameter changes across the transition~\cite{mcwhan70}. The $x$
and $T$ equivalence is also borne out by the observation from XAS at
the $L$-edges in doped V$_2$O$_3$~\cite{park00} that the $a_{1g}$
occupation within both the PM or PI phases is mostly independent of
the doping level. Hence, the local incoherent excitations probed by
XAS at the V $L$ edge or $K$ pre-edge are not directly affected by
disorder~\cite{frenkel06}. The reason for this is that XAS is
a \emph{local probe} in the sense that we can expect the changes in
the XAS spectrum to be of the order of the percentage of the
atoms which have a different ground state.\\

\section{Conclusion}
\label{Sec:conclusion}

The presented findings clearly shows the limits of the common assumption that 
temperature, doping, and pressure--driven MITs in V$_2$O$_3$ can be 
equivalently described within the same phase 
diagram\footnote{An early version of the phase
  diagram (Fig.15 in \cite{mcwhan70}) was actually drawn with a
  third pressure axis, but due to the idea of p--x equivalence, this
  was later abandoned.} \cite{mcwhan70}. Indeed, the two different 
PM electronic structures that we observed reflect different mechanisms 
driving the MIT along different pathways. In the 
doping--driven MIT, the metallic phase is characterized by an increased 
occupation of the $a_{1g}$ orbitals indicating a reduced ``effective 
crystal-field-splitting'' as the main driving mechanism towards 
metallicity~\cite{keller04,poteryaev07}, related to  
the jump of the lattice parameter $c/a$ 
(1.4\%) at the MIT~\cite{mcwhan70}. In contrast, when pressure is 
applied, the $a_{1g}$ occupation remains basically unchanged, so that 
this metallic phase seems to originate rather from an increased 
bandwidth, without any relevant changes of the orbital splitting. The 
smaller $c/a$ jump observed under pressure (0.7 \%) corroborates our 
analysis.

 Doping, temperature and pressure are shown to act
differently on the interplay between electron correlations and
crystal field, so that states previously considered to be equivalent
metals are actually different.

A second important  recent finding is the phase separation within the  ``paramagnetic metallic'' phase for slightly Cr-doped V$_2$O$_3$.
Photoemssion microscopy clearly reveals this phase separation 
 on the microscale, showing that the  ``paramagnetic metallic'' phase is actually a mixture of metallic and insulating regions. This requires the electrons to percolate inbetween the insulating regions, and hence 
leads to a reduced conductivity. In the optical conductivity this is reflected by a pronounced dip at low frequencies. This optical conductivity dip of phase-separated (Cr$_{0.011}$V$_{0.989}$)$_2$O$_3$ can be well described within the effective medium theory, 
either based on the experimental or on the LDA+DMFT optical  conductivities (insulating plus metallic phase).

{\bf Acknowledgement}
We thank O. K. Andersen, M. Capone, M. Haverkort,  N. Parragh and P. Wissgott for valuable discussions regarding
the theory.
The experimental results described in this paper have been obtained
thanks to the collaborative effort of many colleagues, in particular B.
Mansart, L. Baldassarre, F. Rodolakis, E. Papalazarou, A. Perucchi, D.
Nicoletti, J.-P. Rueff, A. Barinov, P. Dudin and L.Petaccia.
We acknowledges financial support from DFG Research Unit FOR 1346
project ID  I597-N16 of  the Austrian Science Fund (FWF),  the EU-Indian network MONAMI, and 
the RTRA Triangle de la
Physique.
The calculations for the results presented 
have been done in part on the Vienna Scientific Cluster (VSC).
\providecommand{\WileyBibTextsc}{}
\let\textsc\WileyBibTextsc
\providecommand{\othercit}{}
\providecommand{\jr}[1]{#1}
\providecommand{\etal}{~et~al.}


\end{document}